# Microwave photonic filters via radio frequency bandwidth scaling with soliton crystal optical micro-combs

Xingyuan Xu,[1] Mengxi Tan,[1] Jiayang Wu,[1] Thach G. Nguyen,[2] Sai T. Chu,[3] Brent E. Little,[4] Roberto Morandotti,[5,6,7] Arnan Mitchell,[2] and David J. Moss[1,a]

[1]*Centre for Micro-Photonics, Swinburne University of Technology, Hawthorn, VIC 3122, Australia*

[2]*School of Engineering, RMIT University, Melbourne, VIC 3001, Australia*

[3]*Department of Physics and Material Science, City University of Hong Kong, Tat Chee Avenue, Hong Kong, China.*

[4]*Xi'an Institute of Optics and Precision Mechanics Precision Mechanics of CAS, Xi'an, China.*

[5]*INRS-Énergie, Matériaux et Télécommunications, 1650 Boulevard Lionel-Boulet, Varennes, Québec, J3X 1S2, Canada.*

[6]*ITMO University, St. Petersburg, Russia.*

[7]*Institute of Fundamental and Frontier Sciences, University of Electronic Science and Technology of China, Chengdu 610054, China.*

We demonstrate high-resolution photonic RF filters using an RF bandwidth scaling approach based on integrated Kerr optical micro-combs. By employing both an active nonlinear micro-ring resonator (MRR) as a high-quality micro-comb source and a passive high-Q MRR to slice the shaped comb, a large RF instantaneous bandwidth of 4.64 GHz and a high resolution of 117 MHz are achieved, together with a broad RF operation band covering 3.28 to 19.4 GHz (L to Ku bands) using thermal tuning. We achieve programmable RF transfer functions including binary-coded notch filters and RF equalizing filters with reconfigurable slopes. Our approach is an attractive solution for high performance RF spectral shaping with high performance and flexibility.

## I. INTRODUCTION

Radio frequency (RF) filters are one of the most basic and commonly used components in modern radar and satellite communications systems [1-5]. While electronics is subject to bandwidth bottlenecks [6], photonic techniques can realize high performance RF filters, offering not only wide bandwidths, but also low loss and strong immunity to electromagnetic interference.

Several approaches have been used for photonic RF filters, including mapping the optical filters' responses onto the RF domain, highlighted by on-chip (waveguide based) stimulated Brillouin scattering [7-9]. This approach has achieved extremely high performance in terms of RF resolution — as high as 32 MHz — and a stopband rejection >55 dB, although it faces challenges in achieving highly reconfigurable transfer functions. Another key approach is based on photonic transversal structures [10-15] that can achieve flexible and arbitrary RF transfer functions simply by changing the tap weights. However, this solution is subject to the tradeoff between resolution and operation bandwidth — for a fixed number of taps, or wavelengths, the product of the two is fixed [6, 16].

Alternative schemes have been demonstrated to achieve reconfigurable RF filters with both high resolution and broad operation bandwidth, including [16] using the tailored phase response of cascaded micro-ring resonators (MRRs). Although successful, it was necessary to use a large number of MRRs to achieve arbitrary transfer functions with high resolution. Recently, "RF bandwidth scaling" [17], employing a two-step approach, has been proposed. First, the RF spectra are imprinted and stretched in the optical domain, assisted by an optical frequency comb and Vernier comb filter. Then the signal is processed by a Waveshaper and compressed to the original RF bandwidth upon photo-detection. This method has enabled high-resolution RF spectral shaping, where the operation bandwidth increases linearly with the number of wavelength channels. However, the use of electro-optic combs [18-21] introduce drawbacks such as the need of high-frequency RF sources to drive the modulators. This in turn results in a limited number of wavelengths that limits the operational bandwidth.

[a]Author to whom correspondence should be addressed. Electronic mail: dmoss@swin.edu.au

Here, we report a high-resolution programmable photonic RF filter achieved by RF bandwidth scaling, enabled by an integrated Kerr optical micro-comb source cascaded with a passive MRR filter. Kerr micro-comb sources [22-29] — particularly those based on CMOS compatible platforms [30-37] — offer many advantages over traditional multi-wavelength sources, such as the potential to provide a much higher number of wavelengths with a greatly reduced footprint and complexity. Here, we experimentally demonstrate programmable arbitrary transfer functions for RF spectral shaping with an operation bandwidth of 4.64 GHz and a resolution of 117 MHz. This was achieved by using a micro-comb source with a free spectral range (FSR) of ~49 GHz (yielding a broadband frequency comb of 80 wavelengths over the C-band and > 160 over the C/L-bands), in combination with a passive MRR with the same FSR and Q factor of 1.5 million. Further, we found that the RF tuning range (controlled via thermal tuning) varies from 3.28 to 19.4 GHz (L to Ku bands). This approach is attractive for high performance photonic RF systems.

## II. OPERATION PRINCIPLE

Figure 1 shows a schematic of the high-resolution photonic RF filter. The Kerr optical frequency combs were generated in a 49GHz integrated MRR (Q factor of 1.5 million) pumped by a continuous-wave (CW) laser. The latter was amplified by an erbium-doped fibre amplifier, with the polarization adjusted via a polarization controller to optimize the power coupled to the MRR. When the pump wavelength was swept across one of the MRR's resonances with the pump power high enough to provide sufficient parametric gain, optical parametric oscillation occurred, ultimately generating Kerr optical combs with a spacing equal to the FSR of the MRR. The generated micro-comb served as a multi-wavelength source with the power of each comb line adjusted by Waveshapers, so to achieve the designed channel weights for constructing arbitrary transfer functions.

In the wavelength range employed by the photonic RF filter, the optical frequency of the $k_{\text{th}}$ ($k$ = 2, 3, 4, …) comb line is

$$f_{OFC}(k) = f_{OFC}(1) + (k-1)\delta_{OFC} \tag{1.1}$$

where $f_{\text{OFC}}(1)$ is the frequency of the first comb line used on the red side, and $\delta_{\text{OFC}}$ is the spacing between comb lines. The shaped comb lines were then passed through a phase modulator, multicasting the input RF signal onto each comb wavelength which was subsequently spectrally sampled, or sliced, by a passive MRR.

For each channel/wavelength the passive MRR introduced a power differential between the two sidebands denoted by a factor k ($0<k<1$). The optical field of the signal after the microring resonator is then given by:

$$E_{opt} = E_0[J_0(\beta)e^{j\omega_0 t} + J_1(\beta)e^{j(\omega_0-\Omega)t} - \sqrt{k}\cdot J_1(\beta)e^{j(\omega_0+\Omega)t}] \tag{1.2}$$

where $E_0$ is the intensity of the optical carrier, $J_n(\beta)$ denote the first kind of the Bessel function with an order $n$, $\beta$ is the modulation depth, $w_0$ is the carrier's angular frequency, and $\Omega$ is the RF frequency. Then the output RF signal after the photodetector is then,

$$I_{RF} \propto |E_{opt}|^2 \propto (1-\sqrt{k})\cdot\cos\Omega t \tag{1.3}$$

Thus, the power imbalance of the two optical sidebands ($k\neq 1$) induced by the optical notch filter (i.e., the passive MRR's through-port transmission) converted the phase modulation into intensity modulation, enabling detection by the photodetector. In addition, since $k$ is a function of $w_0+\Omega$ and denotes the through port transmission of the high-Q resonance optical notch filter, the passive MRR therefore converts the RF phase to intensity modulation, thus achieving a narrow bandpass RF filter for high-resolution RF spectral channelizing. Alternatively it is possible to use the drop port to directly provide intensity-modulation [17]. However, this requires a separate carrier path for multi-heterodyne detection, which is less compact and less stable than our approach.

The FSR of the passive MRR was slightly different to the micro-comb MRR, and so the channelized RF spectral segments on the different comb wavelengths featured progressively staggered RF frequencies with different weights determined by the comb shaping process. The frequency of the $k_{th}$ channelized RF spectral segment is given by:

$$f_{RF}(k) = f_{PMRR}(k) - f_{OFC}(k) = [f_{PMRR}(1) - f_{OFC}(1)] \tag{1.4}$$

where $f_{PMRR}(k)$ is the centre frequency of the passive MRR's $k_{th}$ notch, $[f_{PMRR}(1) - f_{OFC}(1)]$ is the relative spacing between the first comb line and the first filtering resonance, corresponding to the offset of the channelized RF frequencies. The quantity $(\delta_{OFC} - \delta_{PMRR})$ indicates the mismatch between the comb spacing ($\delta_{OFC}$) and the FSR of the passive MRR ($\delta_{PMRR}$), which corresponds to the channelized RF frequency step between adjacent wavelength channels.

Finally, the weighted RF spectra segments were combined upon photo-detection, thus achieving RF bandwidth scaling for arbitrary RF spectral shaping by employing the designed channel weights. Here, the resolution was determined by the passive MRR's Q factor and the operation bandwidth by the product of the resolution and the number of wavelengths.

## III. EXPERIMENTAL RESULTS

The integrated MRRs were fabricated on a high-index doped silica glass (n = ~1.7 at 1550 nm) platform using CMOS-compatible fabrication processes, featuring ultra-low linear loss (~0.06 $dB\cdot cm^{-1}$), a moderate nonlinear parameter (~233 $W^{-1}\cdot km^{-1}$), and, in particular, negligible nonlinear loss up to extremely high intensities (~25 $GW\cdot cm^{-2}$). Due to the ultra-low loss of our platform, the MRRs feature narrow resonance linewidths, corresponding to a Q factor of 1.5 million. After packaging the MRRs with fiber pigtails, the through-port insertion loss was < 2 dB, assisted by on-chip mode converters. The radii of the MRRs were both designed to be ~ 592 μm, corresponding to FSRs of ~0.4 nm (~49 GHz). To realise a mismatch between the comb spacing ($\delta_{OFC}$) and the FSR of the passive MRR ($\delta_{PMRR}$) for RF spectral channelizing, we used orthogonally polarization modes [38, 39] of the active and passive MRRs (TE mode of the active MRR, TM mode of the passive MRR) since they had slightly different effective indices — a result of the non-symmetric waveguide cross sections (~3 μm × 2 μm) that lead to polarization dependent FSRs.

To generate Kerr micro-combs, the pump power was set at ~30.5 dBm and the wavelength was swept across a TE resonance of the active MRR at ~ 1552.9 nm. When the detuning between the pump wavelength and the MRR's cold resonance became small enough, such that the intra-cavity power reached a threshold value, modulation instability driven oscillation was initiated [27]. As the detuning was changed further, distinctive 'fingerprint' optical spectra were observed (Fig. 2(a)). These spectra are similar to what has been reported from spectral interference between tightly packed solitons in the cavity — so called "soliton crystals" [28, 29, 40]. An abrupt step in the measured intracavity power (Fig. 2(b)) was observed at the point where these spectra appeared, as well as a dramatic reduction in the RF intensity noise (Fig. 2(c)). We note that it was not necessary to specifically achieve a single soliton state in order to obtain high performance— only that the chaotic regime [27] should be avoided. This is important since there is a much wider range of coherent low noise states that are more readily accessible than single soliton states [27].

The soliton crystal comb was then spectrally shaped by two stages of Waveshapers (Finisar, 4000S) in order to enable a larger dynamic range of loss control and higher shaping accuracy than a single stage [41, 42]. The micro-comb was first pre-shaped to reduce the power difference between the comb lines to less than 15 dB, then amplified by an erbium-doped fibre amplifier and accurately shaped by a subsequent Waveshaper according to the designed channel weights. For each Waveshaper, a feedback control path was adopted to increase the accuracy of the comb shaping, where the power in the comb lines was detected by an optical spectrum analyzer and compared with the ideal tap weights in order to generate error signals for calibration.

Then the input RF signal was multi-cast onto each comb line via a phase modulator and fed to the passive MRR for spectral slicing. The TM mode through-port transmission of the passive MRR was employed to perform phase-to-intensity modulation conversion by filtering out the lower sideband and, at the same time, map its high-Q resonances (notches) onto the RF domain for high-resolution RF filtering. The optical transmission spectrum of the passive MRR showed an FSR of ~49 GHz (~0.4 nm, Fig. 3(a)) and 3-dB bandwidth of ~0.84 pm (~105 MHz, Fig. 3(b)). We note here that the measured extinction ratio as shown in Fig. 3 was limited by the resolution of our measuring system (~0.1 pm, or 10MHz). The actual extinction ratio was greater than 20dB, as confirmed by the RF transmission results, measured by a vector network analyser with a resolution of 1 kHz.

To illustrate the principle of operation of our high-resolution RF filter, we plot the flattened microcomb spectrum (TE mode) and the drop-port transmission spectrum of the passive MRR (TM mode, measured by using the ASE spectrum of an optical amplifier) in Fig. 4(a). The zoom-in views show the relative shift between the comb lines and the resonances of the passive MRR across the C band, — i.e., the spectrally sliced RF frequencies $f_{RF}(k)$ — ranging from 2.8 to 8.0 GHz. A linear fit of the extracted peak frequencies $f$opt (Fig. 4(b)) determines the FSRs measured to be 48.965 for the passive MRR (TM mode) and 48.898 GHz for the microcomb (TE mode), respectively. Further, it finds a channelized RF frequency step between adjacent channels ($\delta_{OFC} - \delta_{PMRR}$) of ~ 66.6 MHz per channel.

To independently verify this, we measured the RF transmission (Fig. 5) of our filter at the extreme optical bandwidth limits (1$^{st}$ and 80$^{th}$ comb lines, Fig. 5, inset) using a swept frequency RF source with an RF spectrum analyser. Note that the RF bandwidth of each peak is given by the resolution of our system. The spurious lines were a result of the RF source's relatively large frequency step and did not reflect the device performance. The 4.64 GHz spacing between the two passbands indicates an RF frequency step of 4.64 GHz/80 = 58 MHz, in reasonable agreement with the optical measurements.

Finally, the optical signal was converted back into the RF domain by a high-speed photodetector, where the channelized RF spectral segments were combined upon photo-detection. To evaluate the enhanced operation bandwidth enabled by the micro-comb's large number of wavelength channels, we measured the RF transmission spectra with different number of wavelength channels (with flattened comb lines). The single channel case (Fig. 6, blue line) shows a 3dB bandwidth of 117 ± 25 MHz (the ± 25 MHz error bar arises from the resolution of the Vector Network Analyser), which closely matches the measured linewidth of the passive MRR's resonance and defines the resolution of our RF filter. This approach requires dramatically fewer channels, or wavelengths, to achieve this high resolution, unlike transversal filters [43, 44] that require over 800 wavelengths (taps) to reach a similar resolution (with a Nyquist frequency of 20 GHz). Our RF frequency step (58MHz) was about half of the resolution ($\delta_{OFC} - \delta_{PMRR}$ = 117 MHz), whereas ideally, they should be equal. In principle this can be achieved by lithographic control of the waveguide dimensions to vary the polarization dependent FSRs.

From Figure 6 we see that, as the channel number increased from 20 to 80, the operation bandwidth broadened by a factor of four from 1.17 GHz (Fig. 6, yellow line) to 4.64 GHz (Fig. 6, purple line). In addition, employing C+L band Waveshapers and optical amplifiers would yield 160 wavelength channels, leading to a wide instantaneous bandwidth >15 GHz with a similarly high resolution of 117 MHz. The smaller roll-off rate at the high RF frequency side of the wide bandwidth filter results from spurious phase-to-intensity modulation conversion induced by the passive MRR's phase jump, as well as any gain variation with wavelength of the optical amplifier. To minimize these effects, the passive MRR coupling can be optimized to minimize this phase jump, and the gain profile of the optical amplifier equalized with a gain flattening filter.

To demonstrate arbitrary RF transfer functions, we varied the channel weights and measured the RF transmission spectra with a vector network analyser. Figure 7 shows a reconfigurable notch filter achieved with binary (i.e., either "1" or "0") channel weights. The RF transmission spectra (Fig. 7, blue line) match very well with the shaped comb spectra (Fig. 7, red line). Figure 8 shows distinctive RF transmission spectra of RF equalizing filters with flat, quadratic, positive, and negative slopes, implemented by shaping the comb spectra. The RF transmission spectra agree well with the shaped comb spectra, proving the reconfigurability and resolution of our photonic RF filter. We note that flat-top passive optical filters using cascaded MRRs [45-54] can be employed to reduce the passband ripple. The wavelength dependence of the passive MRR loss, RF modulator response, optical amplifiers, and photodetector across the C-band created errors in the channel weights and hence did affect the RF response of the device. This can be compensated for by re-adjusting the channel weights in the Waveshaper by, for example, using an algorithm-driven feedback control path with error signals calculated directly from the generated RF transmission spectra (versus the optical spectra of the shaped comb).

To achieve broadband RF tunability, we thermally tuned the passive MRR to continuously change the relative spacing between the comb lines and passive MRR resonances ($f_{PMRR}(1)$–$f_{OFC}(1)$), where millisecond thermal response times are achievable [55]. Figure 9 shows the measured RF transmission spectra when the chip temperature was varied from 36°C to 44°C, resulting in a wideband RF tunability of 3.3 to 19.4 GHz. By employing a thermal tuning range equivalent to the FSR of the microcomb (~49GHz) and using multiple-FSR spaced filter resonances, an ultra-wideband RF operation of >100 GHz can be achieved.

## IV. CONCLUSION

We demonstrate a high-resolution photonic RF filter based on RF bandwidth scaling by cascading a Kerr optical micro-combs source with a passive high-Q MRR for spectral slicing. A large RF instantaneous bandwidth of 4.64 GHz is achieved using 80 comb lines over the C-band, with a high resolution of 117 MHz and wide RF tuning range from 3-19 GHz (L-Ku bands) via thermal tuning. We achieve programmable RF transfer functions including bandpass and RF equalizing filters with reconfigurable slopes. This represents an attractive approach towards realizing high performance RF filters for modern photonic RF systems [56-82] for systems based on optical microcombs, [83-98] including quantum photonic circuits. [99-111]

## ACKNOWLEDGMENTS

This work was supported by the Australian Research Council Discovery Projects Program (No. DP150104327). RM acknowledges support by the Natural Sciences and Engineering Research Council of Canada (NSERC) through the Strategic, Discovery and Acceleration Grants Schemes, by the MESI PSR-SIIRI Initiative in Quebec, and by the Canada Research Chair Program. He also acknowledges additional support by the Government of the Russian Federation through the ITMO Fellowship and Professorship Program (grant 074-U 01) and by the 1000 Talents Sichuan Program in China. Brent E. Little was supported by the Strategic Priority Research Program of the Chinese Academy of Sciences, Grant No. XDB24030000.

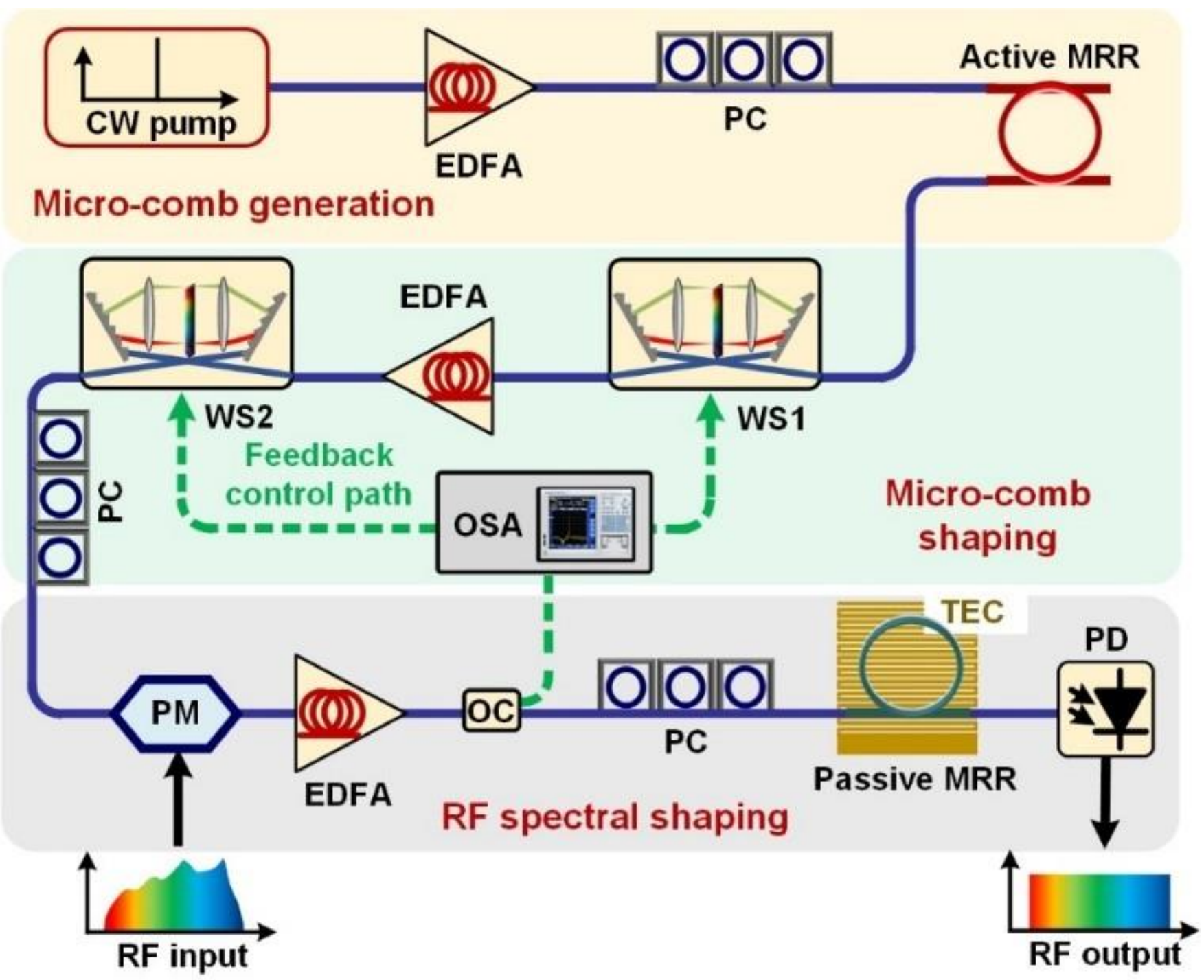

Fig. 1. Schematic diagram of the microcomb-based RF filter. EDFA: erbium-doped fiber amplifier. PC: polarization controller. MRR: micro-ring resonator. WS: Waveshaper. PM: Phase modulator. OC: optical coupler. OSA: optical spectrum analyzer. TEC: Thermoelectric cooler. PD: photodetector.

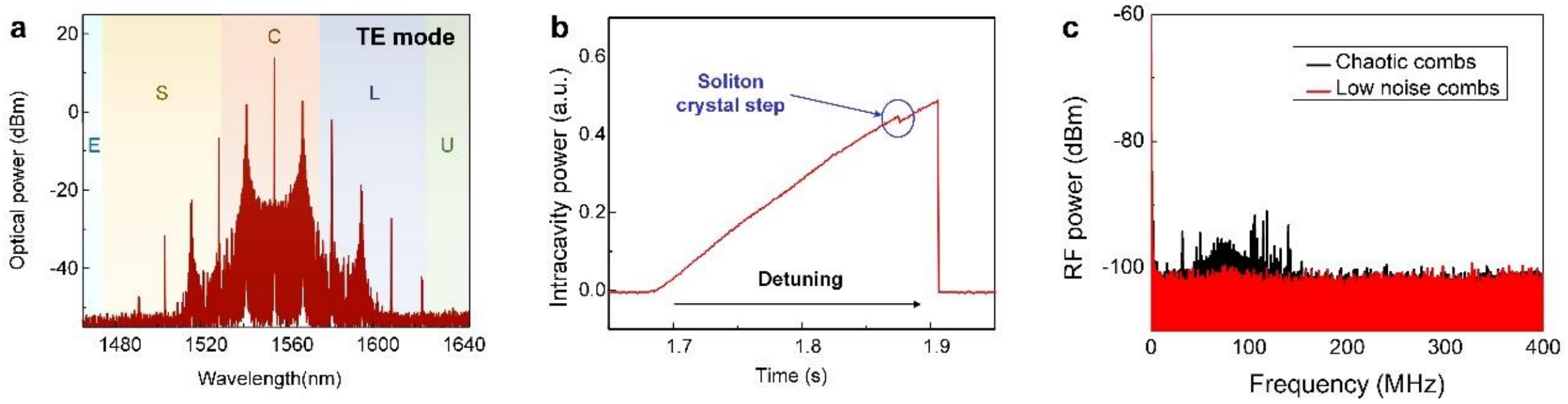

Fig. 2. (a) Optical spectrum of the generated micro-combs with a 200 nm span. (b) Measured transmission of a single resonance showing the soliton crystal step, and (c) the measured RF spectra.

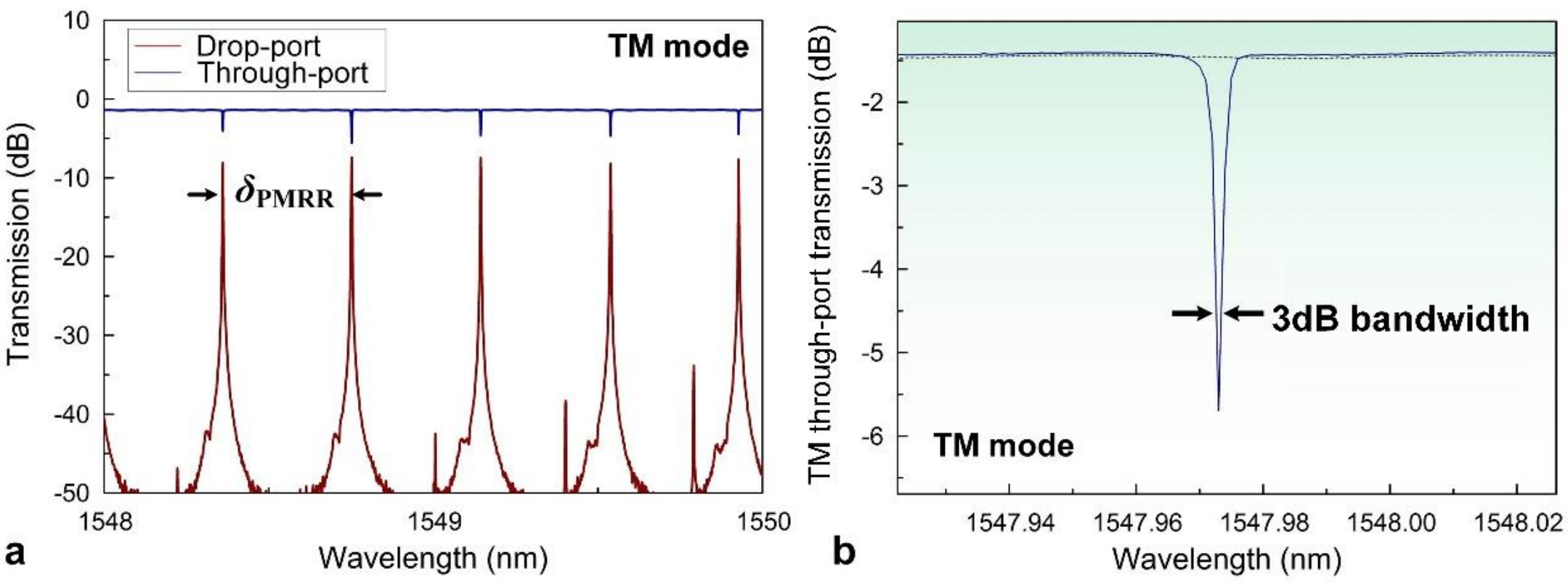

Fig. 3. The TM mode through-port transmission spectra of the passive MRR (a) with a span of 5 nm, (b) showing a resonance at 1547.972 with a 3dB bandwidth of ~0.84 pm, or ~105 MHz. The drop-port was not used but just shown here for completion.

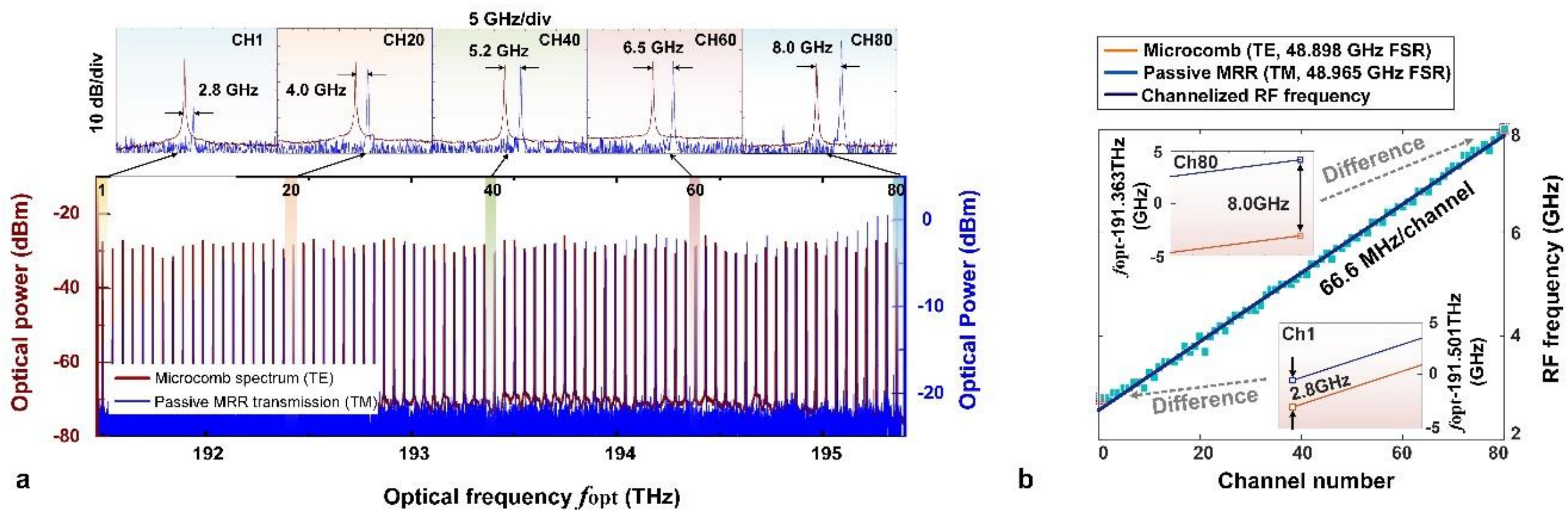

Fig. 4. (a) Measured optical spectrum of the generated micro-comb and transmission spectra of the passive MRR. Zoom-in views of the channels with different channelized RF frequencies. (b) Extracted RF frequencies of the channelized RF spectral segments.

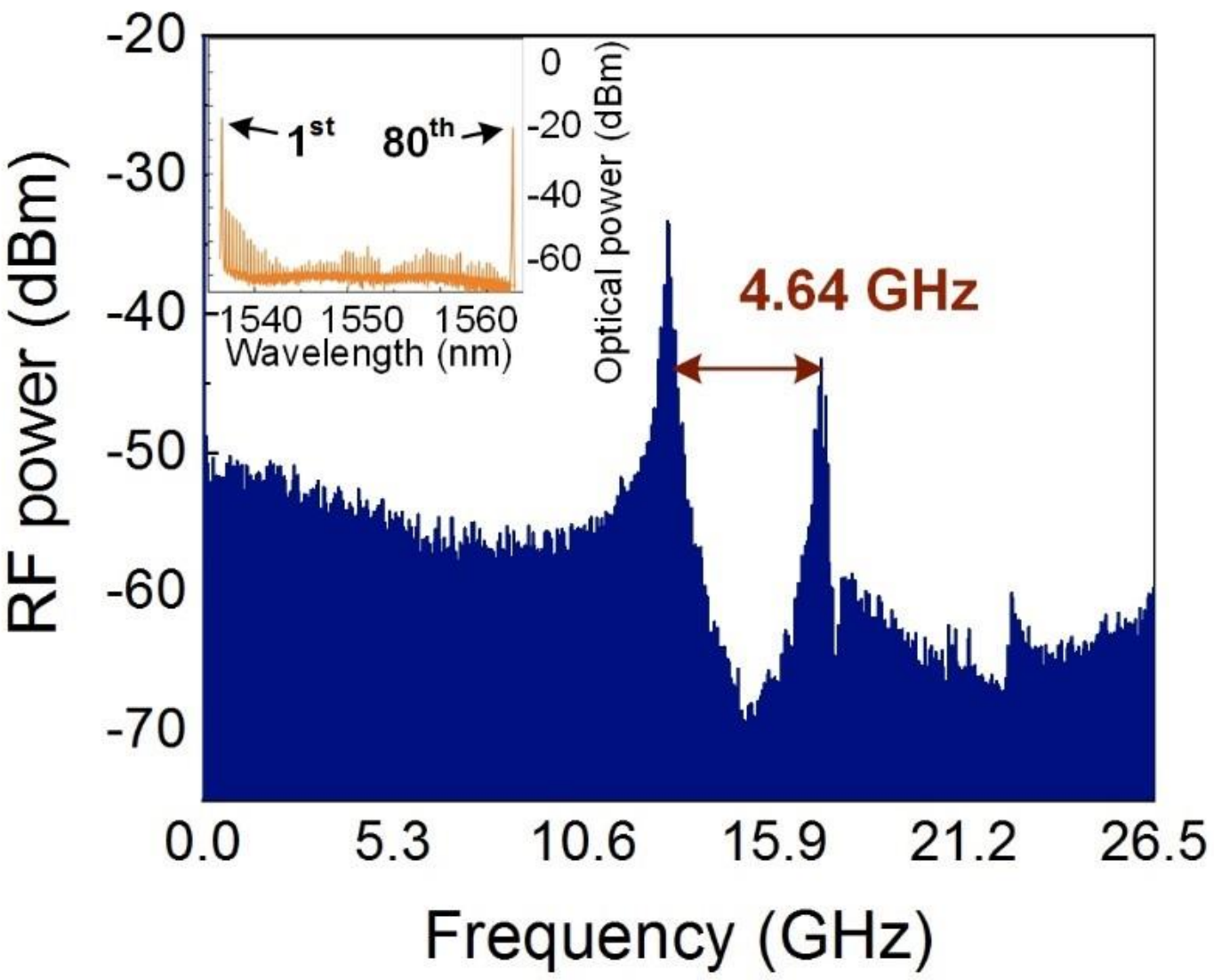

Fig. 5. Measured RF transmission spectra of the RF filter with the 1st and 80th comb line being employed. The inset shows the corresponding optical spectrum.

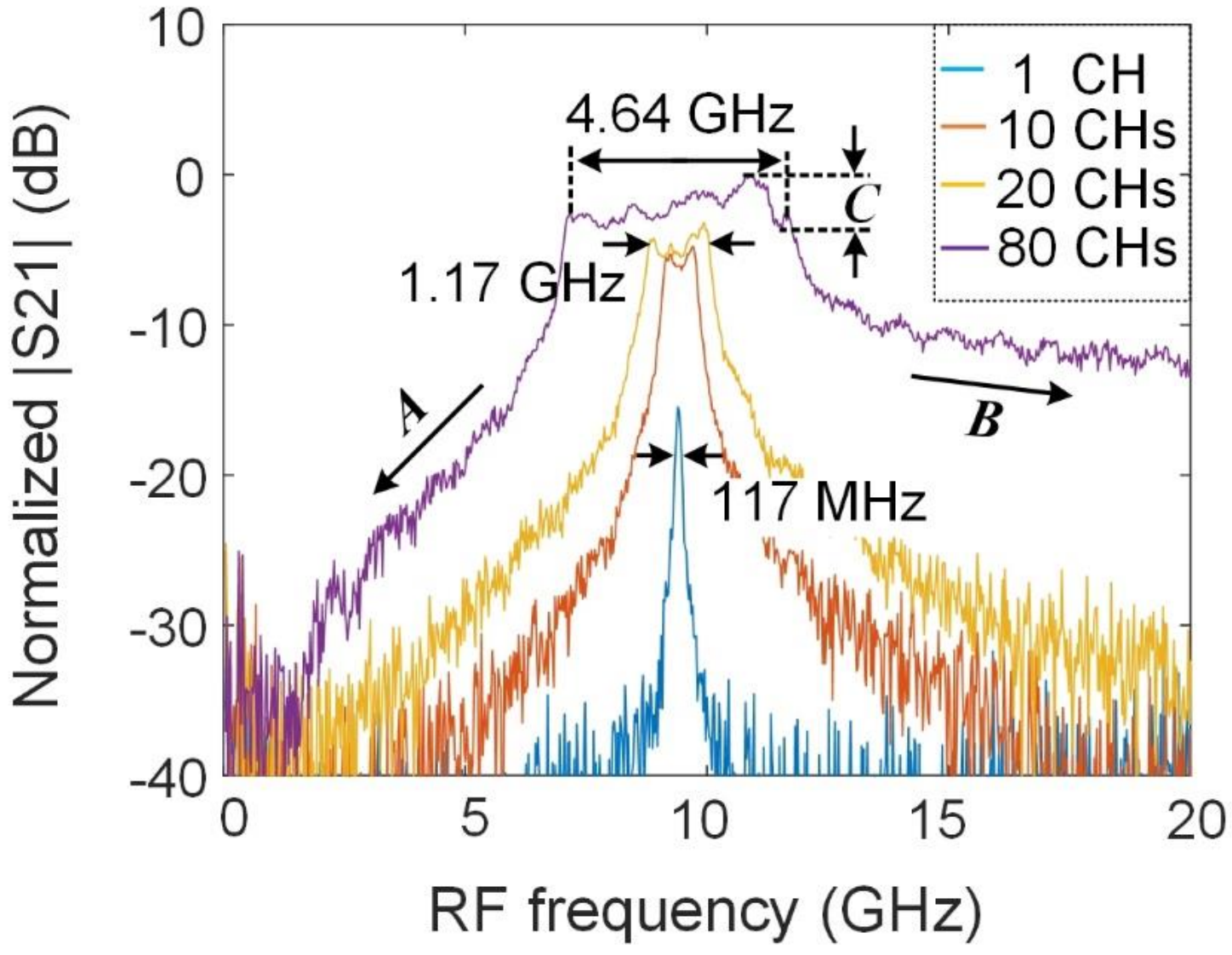

Fig. 6. Measured RF transmission spectra of the RF filter as the number of wavelength channels changed from 1 to 80. A: roll-off rate = 5.2 dB/GHz. B: roll-off rate = 1.3 dB/GHz. C: roll-off rate = 2.9 dB/GHz.

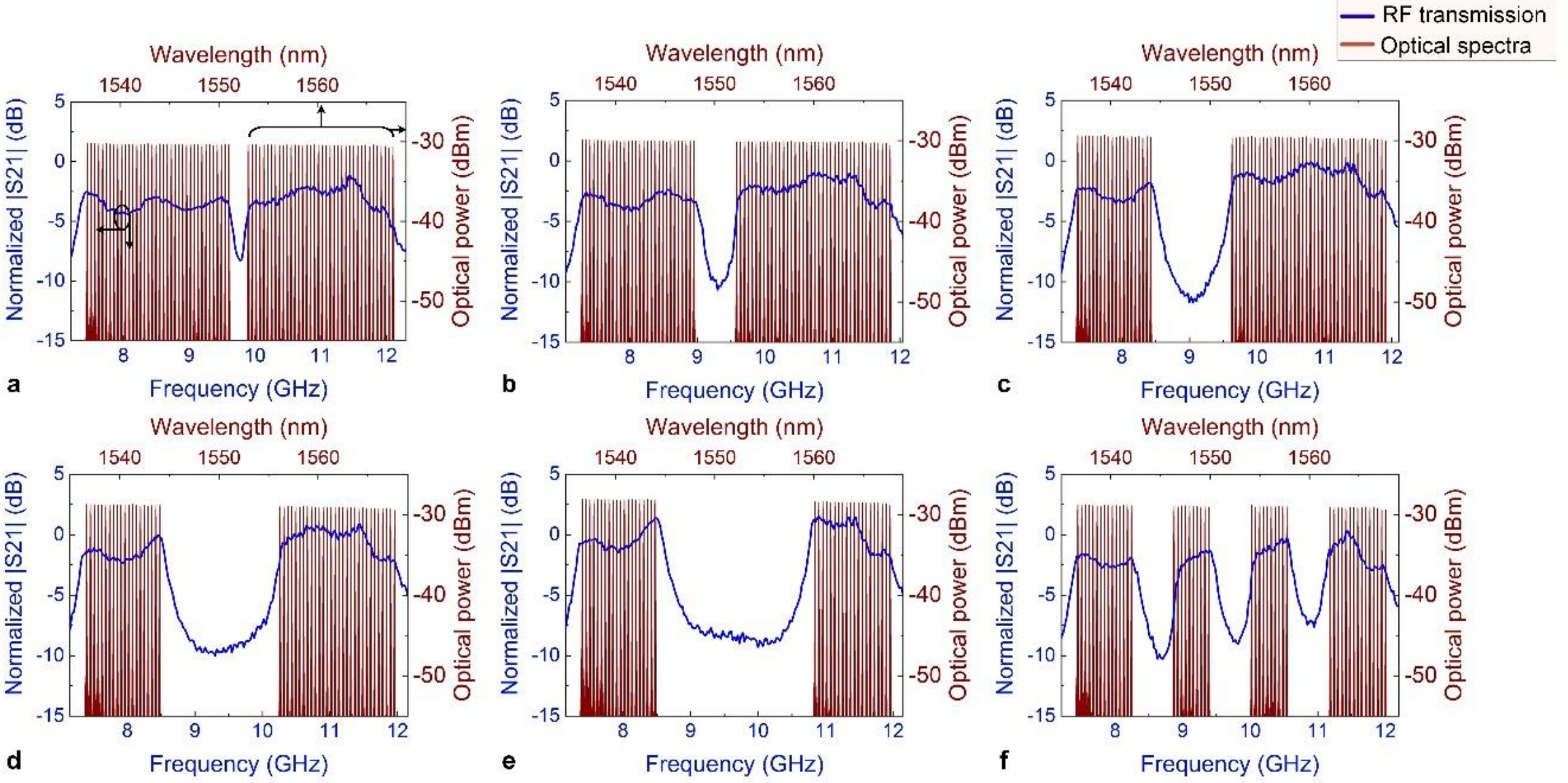

Fig. 7. Measured RF transmission spectra (blue line) of the RF filter featuring binary coded channel weights and corresponding optical spectra (red line) of the shaped micro-comb.

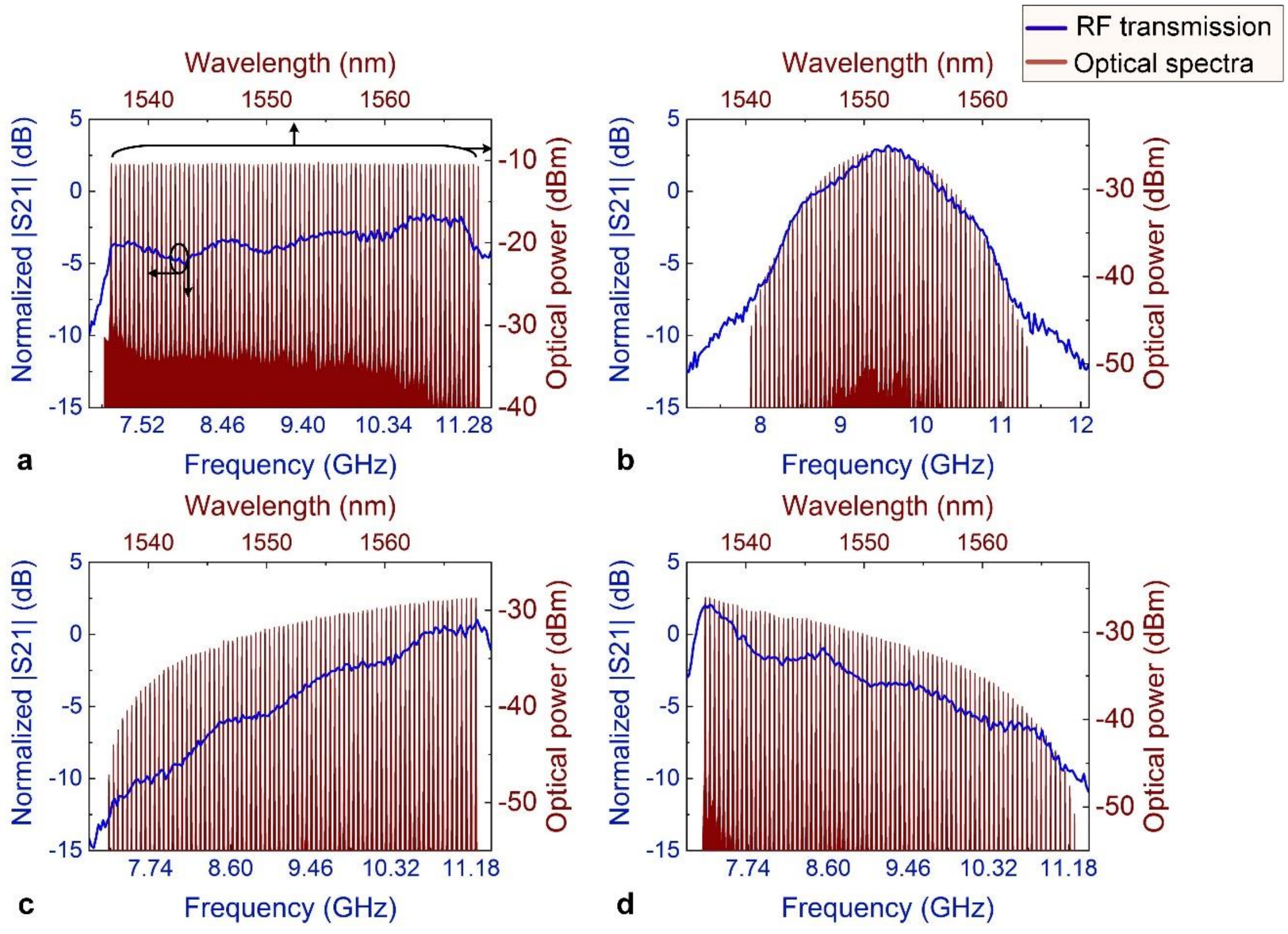

Fig. 8. Measured RF transmission spectra (blue line) of the RF equalizing filter featuring varying slopes and corresponding optical spectra (red line) of the shaped micro-comb.

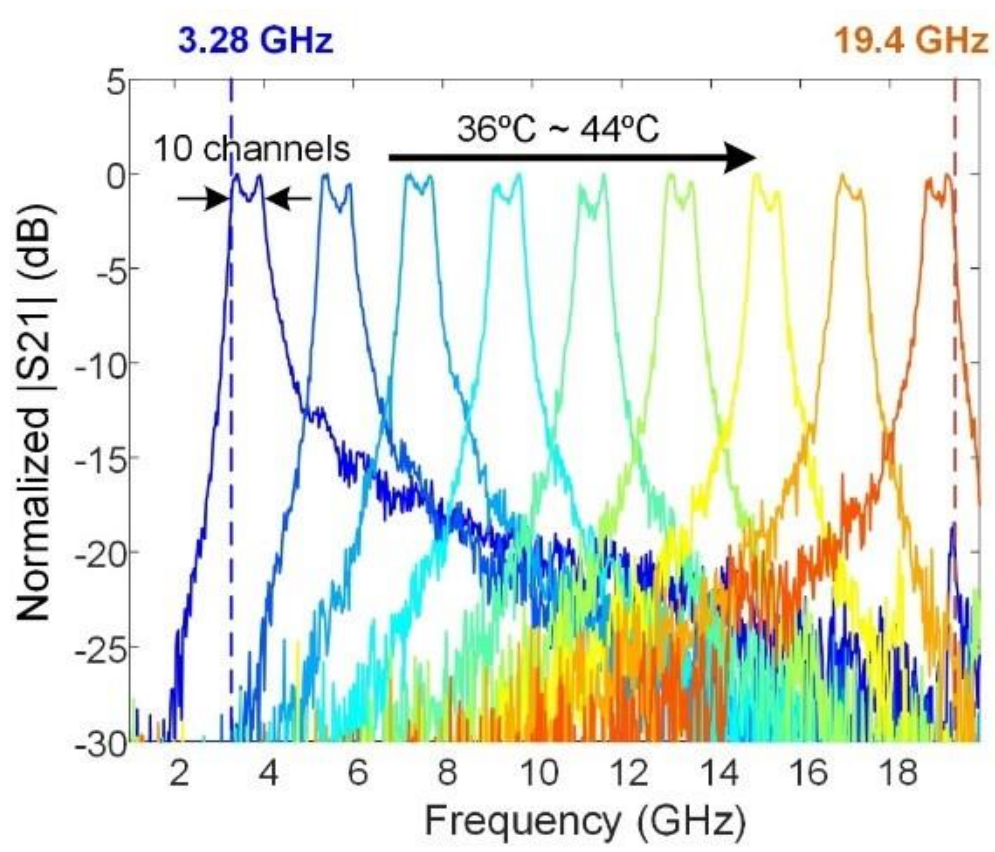

Fig. 9. Measured RF transmission spectra of the RF filter with a varying chip temperature of the passive MRR.